\documentclass{article}   
\usepackage{graphicx}
\input{abarbanel.def}
\begin{document}
\begin{center}
{\Large\bf Reading Neural Encodings using Phase Space Methods} \\
{\large Henry D. I. Abarbanel\dag \ and Evren C. Tumer} \\
Department of Physics and Institute for Nonlinear Science \\
University of California, San Diego \\
email : evren@nye.ucsd.edu \\
March 2003 \\
\dag also Marine Physical Laboratory, Scripps Institute of Oceanography \\
\end{center} 
\begin{it} Dedicated to Larry Sirovich on the occasion of his 70th
birthday
\end{it}

\begin{abstract}

Environmental signals sensed by nervous systems are often represented in
spike trains carried from sensory neurons to higher neural functions where
decisions and functional actions occur. Information about the environmental
stimulus is contained (encoded) in the train of spikes. We show how to
``read'' the encoding using state space methods of nonlinear dynamics. We
create a mapping from spike signals which are output from the neural
processing system back to an estimate of the analog input signal. This
mapping is realized locally in a reconstructed state space embodying both
the dynamics of the source of the sensory signal and the dynamics of the
neural circuit doing the processing. We explore this idea using a
Hodgkin-Huxley conductance based neuron model and input from a low
dimensional dynamical system, the Lorenz system. We show that one may
accurately learn the dynamical input/output connection and estimate with
high precision the details of the input signals from spike timing output
alone.
This form of ``reading the neural code'' has a focus on the neural circuitry
as a dynamical system and emphasizes how one interprets the dynamical
degrees of freedom in the neural circuit as they transform analog
environmental information into spike trains. 
\end{abstract}

\section{Introduction}
A primary task of nervous systems is the collection at its periphery
of information from the environment and the distribution of that
stimulus input to central nervous system functions. This is often
accomplished through the production and transmission of action 
potentials or {\em spike}  trains~\cite{spikes}. 

The book~\cite{spikes} and subsequent papers by its authors and their
collaborators~\cite{synergy} carefully lay out a program for interpreting
the analog stimulus of a nervous system using ideas from probability theory
and information theory, as well as a representation of the input/output or
stimulus/response relation in terms of Volterra kernel functions.
In~\cite{spikes} the authors note that when presenting a stimulus to a
neuron, it is common ``that the response spike train is not identical on
each trial.'' Also they observe that ``Since there is no unique response,
the most we can say is that there is some probability of observing each of
the different possible responses." This viewpoint then underlies the wide
use of probabilistic ideas in describing how one can ``read the neural
code'' through interpreting the response spike trains to infer the stimulus.

In this paper we take a different point of view and recognize that the
neuron into which one sends a stimulus is itself a dynamical system with a
time dependent state which will typically be different upon receipt of
different realizations of identical stimulus inputs. Viewing the
transformation of the stimulus waveform into the
observed response sequence, as a result of deterministic dynamical action of the 
neuron one can attribute the variation in the response
to identical stimuli to differing neuron states when the stimulus
arrives. This allows us to view the entire transduction
process of analog input (stimulus) to spike train output (response) as a
deterministic process which can be addressed by methods developed in
nonlinear dynamics for dealing with input/output systems~\cite{rhodes}. 

Previous research on information encoding in spike trains has concentrated
on nonlinear filters that convert analog input signals into spike trains. 
It has been shown that these models can be used to reconstruct the dynamical
phase space of chaotic inputs to the filters using the spike  timing
information~\cite{sauerISI,castro_sauer2,pavlov,pavlov2}.  Using simple
dynamical neuron models, Castro and Sauer~\cite{castro_sauer} have shown
that aspects of a dynamical system can be reconstructed using interspike
intervals (ISIs) properties.  Experimental work has demonstrated the
ability to discriminate between chaotic and stochastic inputs to a
neuron~\cite{richardson}, as well as showing that decoding sensory
information from a spike train through linear filtering Volterra
series techniques can allow for large amounts of information to be
carried by the precise timing of the spikes~\cite{spikes}.

We discuss here the formulation of input/output systems from a dynamical
system point of view, primarily summarizing earlier work~\cite{rhodes,book},
but with a focus on recognizing that we may treat the response signals as
trains of identical spikes. Since the modulation of the spike train must be
carrying the information in the analog input presented to the neuron, if the
spike pulse shapes are identical, all information must be encoded in the
ISIs. We shall show that this is, indeed, the case.

What is the role of information theory in a deterministic chain of actions
from stimulus to spiking response? The ideas of information theory, though
often couched in terms of random variables, applies directly to distributed
variation in dynamical variables such as the output from nonlinear systems.
The use of concepts such as entropy and mutual information, at the basis of
information theoretic descriptions of systems, applies easily and directly
to deterministic systems. The understanding of this connection dates from
the 1970's and 1980's where the work of Fraser~\cite{fraser} makes this
explicit, and the connection due to Pesin~\cite{pesin} between positive
Lyapunov exponents of a deterministic system and the Kolmogorov-Sinai
entropy quantifies the correspondence.

In the body of this paper, we first summarize the methods used to determine
a connection between analog input signals and spiking output, then we apply
these methods to a Hodgkin-Huxley conductance based model of the R15 neuron 
of {\em Aplysia}~\cite{canavier,plant}. Future papers will
investigate the use of these methods on biological signals from the H1
visual neuron of a fly and a stretch receptor in the tail of a
crayfish~\cite{tumer}

\section{Input Estimation from State Space Reconstruction}

The general problem we address is the response to stimuli of a neural
circuit with N dynamical variables 
$$
\mathbf{x}(t) = [x_1(t),x_2(t),\ldots,x_N(t)].
$$ 
When there is no time varying
input, $\mathbf{x}(t)$ satisfies the ordinary differential equations
\begin{equation}
\frac{d x_a(t)}{dt} = F_a(\mathbf{x}(t)), \,\,\,; a = 1, 2, ..., N
\end{equation} 
The $F_a(\mathbf{x})$ are a set of nonlinear functions which determine the
dynamical time course of the neural circuit. The $F_a(\mathbf{x})$ could well
represent a conductance based neural model of the Hodgkin-Huxley variety as
in our example below.

When there is a time dependent external stimulus $s(t)$, these equations
become
\begin{equation}
\frac{d x_a(t)}{dt} = F_a(\mathbf{x}(t),s(t)), 
\label{nonautoDS}
\end{equation} 
 and the time course of $\mathbf{x}(t)$ in this driven or non-autonomous setting
can become rather more complicated than the case where $s(t) =
\mbox{constant}$.

If we knew the dynamical origin of the signal $s(t)$, then in the combined
space of the stimuli and the neural state space $\mathbf{x}(t)$, we would again have
an autonomous system, and many familiar~\cite{book} methods for analyzing
signals from nonlinear systems would apply.  As we proceed to our ``input
signal from spike outputs'' connection we imagine that the stimulus system
is determined by some other set of state variables $\z(t)$ and that
\bea
\frac{d \z(t)}{dt} &=& \G(\z(t)) \nonumber \\ s(t) &=& h(\z(t)),
\eea
where $\G(\z)$ are the nonlinear functions determining the time course of
the state $\z(t)$ and $h(\z(t))$ is the nonlinear function determining the
input to the neuron $s(t)$.

With observations of just one component of the state vector $\mathbf{x}(t)$, the
full dynamical structure of a system described by Equation~\ref{nonautoDS}
can be reconstructed in a proxy state space~\cite{Mane,Takens}.  Once the
dynamics of the system is reconstructed, the mapping from state variable to
input can be made in the reconstructed space.  Assume the measured state
variable, $r(t) = g(\x(t))$, is sampled at times $t_j$, where $j$ is an
integer index.  According to the embedding theorem~\cite{Mane,Takens},
the dynamics of the system can be reconstructed in an embedding space
using time delayed vectors of the form
\bea
 \label{tdev}
\y(j) &=& [r(t_j) , r(t_j + T\tau_s) , \ldots , r(t_j + (\de-1)T\tau_s)]
\nonumber \\ &=& [r(j),r(j+T), \dots,r(j+(\de-1)T)]
\eea
where $\de$ is the dimension of the embedding, $t_j = t_0 + j\tau_s$,
$\tau_s$ is the sampling time, $t_0$ is an initial time, and $T$
is an integer time delay. If the dimension $\de$ is large enough
these vectors can reconstruct the dynamical structure of the full system
given in Equation~\ref{nonautoDS}. Each vector $\y(j)$ in the reconstructed
phase space depends on the state of the input signal.  Therefore a mapping
should exist that associates locations in the reconstructed phase space
$\y(j)$ to values of the input signal $s(t_j) \equiv s(j): s(j) = H(\y(j))$.
The map $H(\y)$ is the output-to-input relation we seek.

Without simultaneous measurements of the observable $r(t)$ and the input
signal $s(t)$, this mapping could not be found without knowing the
differential equations that make up Equation~\ref{nonautoDS}.  But in a
situation where a controlled stimulus is presented to a neuron while
measuring the output, both $r(t)$ and $s(t)$ are available
simultaneously.  Such a data set with simultaneous measurements of
spike time and input is split into two parts: the first part, called the
training set, will be used to find the mapping $H(\y(j))$ between 
$\y(j)$ and $s(j)$.  The second part, called the test set, will
be used to test the accuracy of that mapping.
State variable data from the training set $r(j)$ is used to
construct time delayed vectors as given by
\begin{equation}
\label{train_vec}
\y(j) = [ r(j), r(j+T), \ldots , r(j + (\de-1)T)].
\end{equation} Each of these vectors is paired with the value of the
stimulus at the midpoint time of the delay vector
\begin{equation}
\label{train_val} s(j) = s\left(t_{j + T(\de-1)/2}\right)
\end{equation}
We use state space values that occur before and after the
input to improve the quality of the representation.  The state variables and
input values in the remainder of the data are organized in a similar way
and used to test the mapping.

The phase space dynamics near a test data vector are reconstructed using
vectors in the training set that are close to the test vector, where we use
Euclidian distance between vectors. These vectors lie close in the
reconstructed phase space, so they will define the dynamics of the system in
that region and will define a {\bf local} map from that region to a input
signal value.  In other words, we seek a form for $H(\y(j))$ which is local
in reconstructed phase space to $\y(j)$. The {\bf global} map over all of
phase space is a collection of local maps.

The local map is made using the $N_B$ nearest neighbors $\y^{m}(j)\;,\;m=0
\ldots N_B$ of $\y^0(j) = \y(j)$.  These nearest neighbor vectors and
their corresponding input values $s^{m}(j)$ are used to find a local
polynomial mapping between inputs $s^m(j)$ and vector versions of the
outputs $r^m(j)$, namely $\y^m(j)$ of the form
\begin{equation}\label{localmap} 
s^m(j) = H(\y^m(j)) = M_0(j) + \M_1(j) \cdot \y^m(j) + \M_2(j) \cdot \y^m(j)
\cdot \y^m(j) + \cdots,
\end{equation} 
which assume that the function $H(\y)$ is locally smooth in phase space.

The scalar $M_0(j)$, the $\de$-dimensional vector $\M_1(j)$, and the tensor
$\M_2(j)$ in $\de$-dimensions, etc are determined by minimizing the mean
squared error 
\begin{equation} \label{ls_error}
\sum^{N_B}_{m=0} |s^m(j) - M_0(j) + \M_1(j) \cdot \y^m(j) + \M_2(j) \cdot
\y^m(j) \cdot \y^m(j)  + \cdots |^2.
\end{equation}
We determine $M_0(j), \M_1(j), \M_2(j), \ldots$ for all  $j = 1,
2,\ldots$, and this provides a local representation of $H(\y)$ in all parts
of phase space sampled by the training set
$\y(j), \; j = 1, 2,\ldots,N_{train}$.

Once the least squares fit values of $M_0(j), \M_1(j), \M_2(j),\ldots$ are
determined for our training set, we can use the resulting local map to
determine estimates of the input associated with an observed output. This
proceeds as follows: select a new output $r^{new}(l)$ and form the new
output vector $\y^{new}(l)$ as above. Find the nearest neighbor in the
training set to $\y^{new}(l)$. Suppose it is the vector $\y(q)$. Now
evaluate an estimated input $s^{est}(l)$ as
\begin{equation} 
s^{est}(l) = M_0(q) + \M_1(q) \cdot \y^{new}(l) + \M_2(q) \cdot
\y^{new}(l) \cdot
\y^{new}(l) + \cdots.
\end{equation} 
This procedure is applied for all new outputs to produce the
corresponding estimated inputs.

\section{R15 Neuron Model}

To investigate our ability to reconstruct stimuli of analog form presented
to a realistic neuron from the spike train output of that neuron, we
examined a detailed model of the R15 neuron in
{\em Aplysia}~\cite{canavier,plant}, and presented this model neuron
with nonperiodic input from a  low dimensional dynamical system. This model
has seven dynamical degrees of freedom. The differential equations for this
model are
\bea C\frac{d V_m(t)}{dt} &=& (g_Iy_2(t)^3y_3(t) + g_T)(V_I-V(t))
+ g_L(V_L-V(t))
\nonumber \\ &+& (g_Ky_4(t)^4 + g_Ay_5(t)y_6(t)+g_Py_7(t))(V_K-V(t))
\nonumber \\ &+& I_0 + I_{ext} + I_{input}(t),
\eea
where the $y_n(t); \; n =2,3,\ldots,7$ satisfy kinetic equations of the form
\begin{equation} 
\frac{dy_n(t)}{dt} = \frac{Y_n(V_m(t))-y_n(t)}{\tau_n(V_m(t))},
\end{equation} 
which is the usual form of Hodgkin-Huxley models. The $g_X, X = I, T, K,
A, P, L$ are maximal conductances, the $V_X, X = I, L, K$ are reversal
potentials. $V_m(t)$ is the membrane potential, $C$ is the membrane
capacitance, $I_0$ is a fixed DC current,  and $I_{ext}$ is a DC current we
vary to change the state of oscillation of the model. The functions $Y_n(V)$
and $\tau_n(V)$ and values for the various constants are given
in~\cite{canavier,plant}. These are phenomenological forms of membrane voltage
dependent gating variables, activation and inactivation of membrane ionic
channels, and time constants for these gates. $I_{input}(t)$ is a time
varying current input to the neural dynamics. Our goal will be to reconstruct
$I_{input}(t)$ from observations of the spike timing in $V_m(t)$.

\begin{figure}[ht]
\begin{center}
\includegraphics[width=9cm,height=9cm,angle=-90]{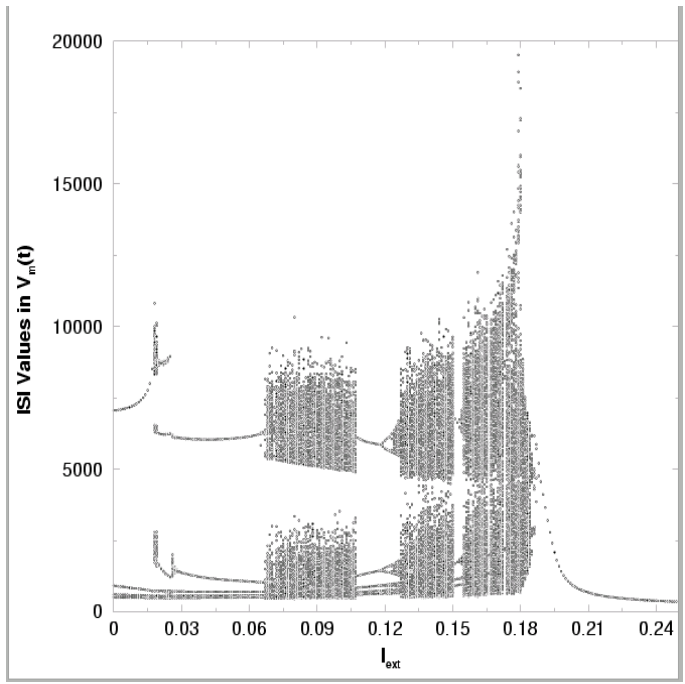}
\caption{Bifurcation diagram for the R15 model with constant input current.
This plot shows the values of ISIs which occur in the $V_m(t)$ time series
for different values of $I_{ext}$.}
\label{bifur_diagram}
\end{center}
\end{figure}

In Figure~\ref{bifur_diagram} we plot the bifurcation diagram of our R15
model. On the vertical axis we show the values of ISIs taken in the time
series for $V_m(t)$ from the model; on the horizontal axis we plot
$I_{ext}$. From this bifurcation plot we see that the output of the R15
model has regular windows for $I_{ext} < .07$ then chaotic regions
interspersed with periodic orbits until $I_{ext} \approx 0.19$ after which
nearly periodic behavior is seen. The last region represents significant
depolarization of the neuron in which tonic periodic firing associated
with a stable limit cycle in phase space is typical of neural activity.
Periodic firing leads to a fixed value for ISIs, which is what we see.
Careful inspection of the time series reveals very small fluctuations in the
phase space orbit, but the resolution in Figure~\ref{bifur_diagram} does not
expose this.

Other than the characteristic spikes, there are no significant features in
the membrane voltage dynamics. In addition all the spikes are essentially
the same, so we expect that all the information about the membrane voltage
state is captured in the times between spikes, namely the interspike
intervals: ISIs. The distribution of ISIs characterizes the output signal
for information theoretic purposes.

We have chosen three values of $I_{ext}$ at which to examine the response of
this neuron model when presented with an input signal. At $I_{ext} = 0.1613$
we expect chaotic waveforms expressed as nonperiodic ISIs with a broad
distribution. At $I_{ext} = 0.2031$ we expect nearly periodic spike trains.
And at $I_{ext} = -0.15$ the neuron does not spike, the mebrane voltage remains
at an equilibrium value.

For each $V_m(t)$ time series we evaluate the normalized distribution of
ISIs which we call $P_{ISI}(\Delta)$ and from this we compute the entropy
associated with the oscillations of the neuron. Entropy is defined as
\begin{equation} 
H(\Delta) = \sum_{\mbox{observed}\, \Delta} - P_{ISI}(\Delta) 
\log \biggl( P_{ISI}(\Delta)
\biggr);
\end{equation} 
$H(\Delta) \ge 0$. The entropy is a
quantitative measure~\cite{shannon} of the information content of
the output signal from the neural activity.

\begin{figure}[ht] 
\begin{center}
\includegraphics[width=9cm,angle=0]{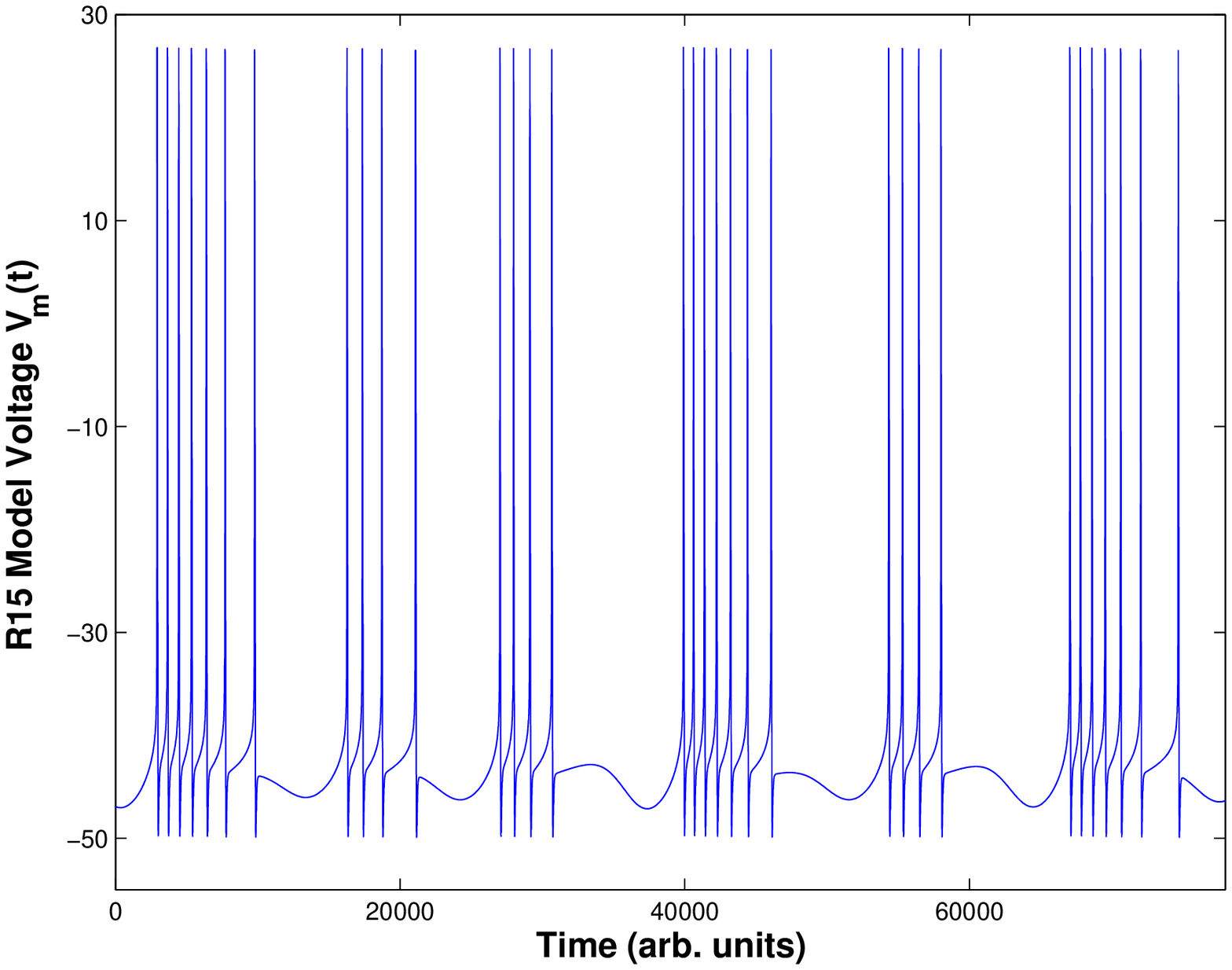}
\caption{Membrane voltage of the R15 model with a constant input current
$I_{ext}$ = 0.1613.}
\label{spikes1613}
\end{center}
\end{figure}

\begin{figure}[ht] 
\begin{center}
\includegraphics[width=9cm,height=9cm,angle=0]{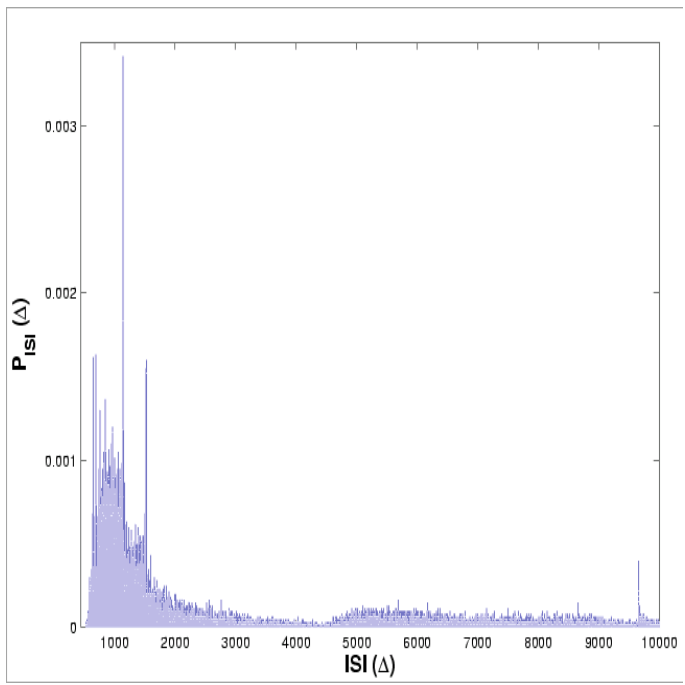}
\caption{Normalized distribution $P_{ISI}(\Delta)$ from the membrane voltage
time series with $I_{ext}$ = 0.1613.  The entropy for this distribution
$H(\Delta) = 12$.}
\label{isidist1613}
\end{center}
\end{figure}

\begin{figure}[ht] 
\begin{center}
\includegraphics[width=9cm,height=9cm,angle=0]{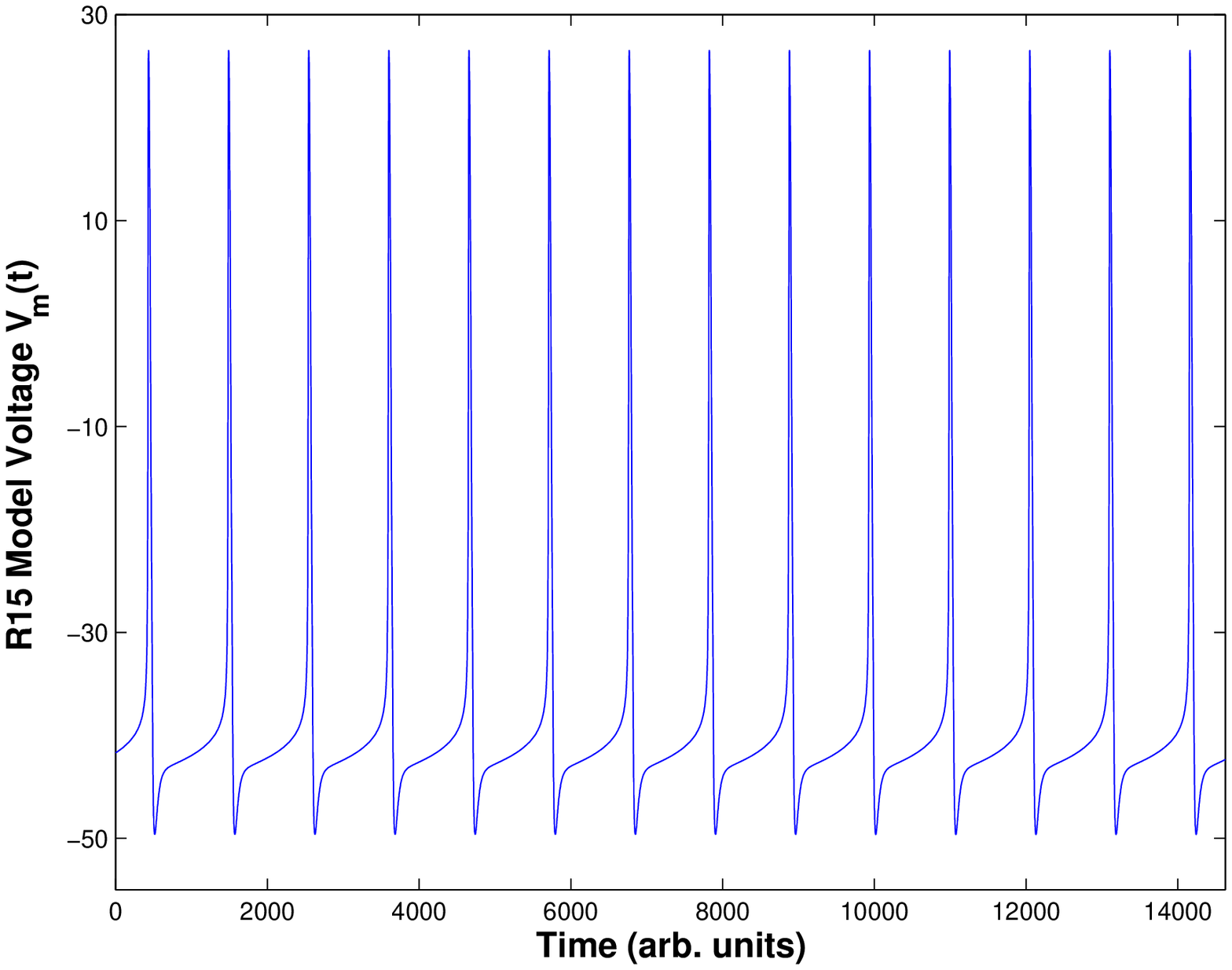}
\caption{Membrane voltage of the R15 model with a constant input current
$I_{ext}$ = 0.2031.}
\label{spikes2011}
\end{center}
\end{figure}

In Figure~\ref{spikes1613} we display a section of the $V_m(t)$ time series
for $I_{ext} = 0.1613$. The irregularity in the spiking times is clear from
this figure and the distribution $P_{ISI}(\Delta)$ shown in
Figure~\ref{isidist1613}.  The $P_{ISI}(\Delta)$ was evaluated from
collecting 60,000 spikes from the $V_m(t)$ time series and creating
a histogram with 15,000 bins. This distribution has an entropy
$H(\Delta) = 12$.  In contrast to this we have a section of the
$V_m(t)$ time series for $I_{ext} = 0.2031$ in Figure~\ref{spikes2011}.
Far more regular firing is observed with a firing frequency much
higher than for $I_{ext} = 0.1613$. This increase in firing
frequency as a neuron is depolarized is familiar. With $I_{ext} = 0.2031$ 
the distribution $P_{ISI}(\Delta)$ is mainly concentrated in one bin
with some small flucuations near that bin.  Such a regular distribution
leads to a very low entropy $H(\Delta) = 0.034$.  If not for the
slight variations in ISI, the entropy would be zero.  If
$P_{ISI}(\Delta_0) = 1$ for some ISI value $\Delta_0$, then
$H(\Delta) = 0$.

\subsection{Input Signals to Model Neuron}

In the last section the dynamics of the neuron model were examined using
constant input signals. In studying how neurons encode information in their
spike train, we must clarify what it means for a signal to carry
information.  In the context of information theory~\cite{shannon},
information lies in the unpredictability of a signal.  If we do not know
what a signal is going to do next, then by observing it we gain new
information. Stochastic signals are commonly used as information carrying
signals since their unpredictability is easily characterized and readily
incorporated into the theoretical structure of information theory. But they
are problematic when approaching a problem from a dynamical systems point of
view, since they are systems with a high dimension. This means that the
reconstruction of a stochastic signal using time delay embedding vectors
of the form of Equation~\ref{tdev} would require an extremely large embedding
dimension~\cite{book}. If we are injecting stochastic signals into the R15
model, the dimension of the whole system would increase and cause practical
problems in performing the input reconstruction.  Indeed, the degrees of
freedom in the stochastic input signal could well make the input/output
relationship we seek to expose impossible to see.

An attractive input for testing the reconstruction method will have some
unpredictability but have few degrees of freedom. If there are many degrees
of freedom, the dimensionality of the vector of outputs $\y(j)$ above may be
prohibitively large.  This leads directly to the consideration of low
dimensional chaotic systems.  Chaos originates from local instabilities
which cause two points initially close together in phase space to
diverge rapidly as the system evolves in time, thus producing
completely different trajectories.  This exponential divergence is
quantified by the positive Lyapunov exponents and is the source of the
unpredictability in chaotic systems~\cite{book}. The state of any observed
system is known only to some degree of accuracy, limited by measurement and
systematic errors.  If the state of a chaotic system were known exactly then
the future state of that system should be exactly predictable.  But if the
state of a chaotic system is only known to some finite accuracy, then
predictions into the future based on the estimated state will diverge from
the actual evolution of the system.  Imperfect observations of a chaotic
signal will limit the predictability of the signal.   Since chaos can occur
in low dimensional systems these signals do not raise the same concerns as
stochastic signals.

\begin{figure}[ht] 
\begin{center}
\includegraphics[width=9cm,height=9cm]{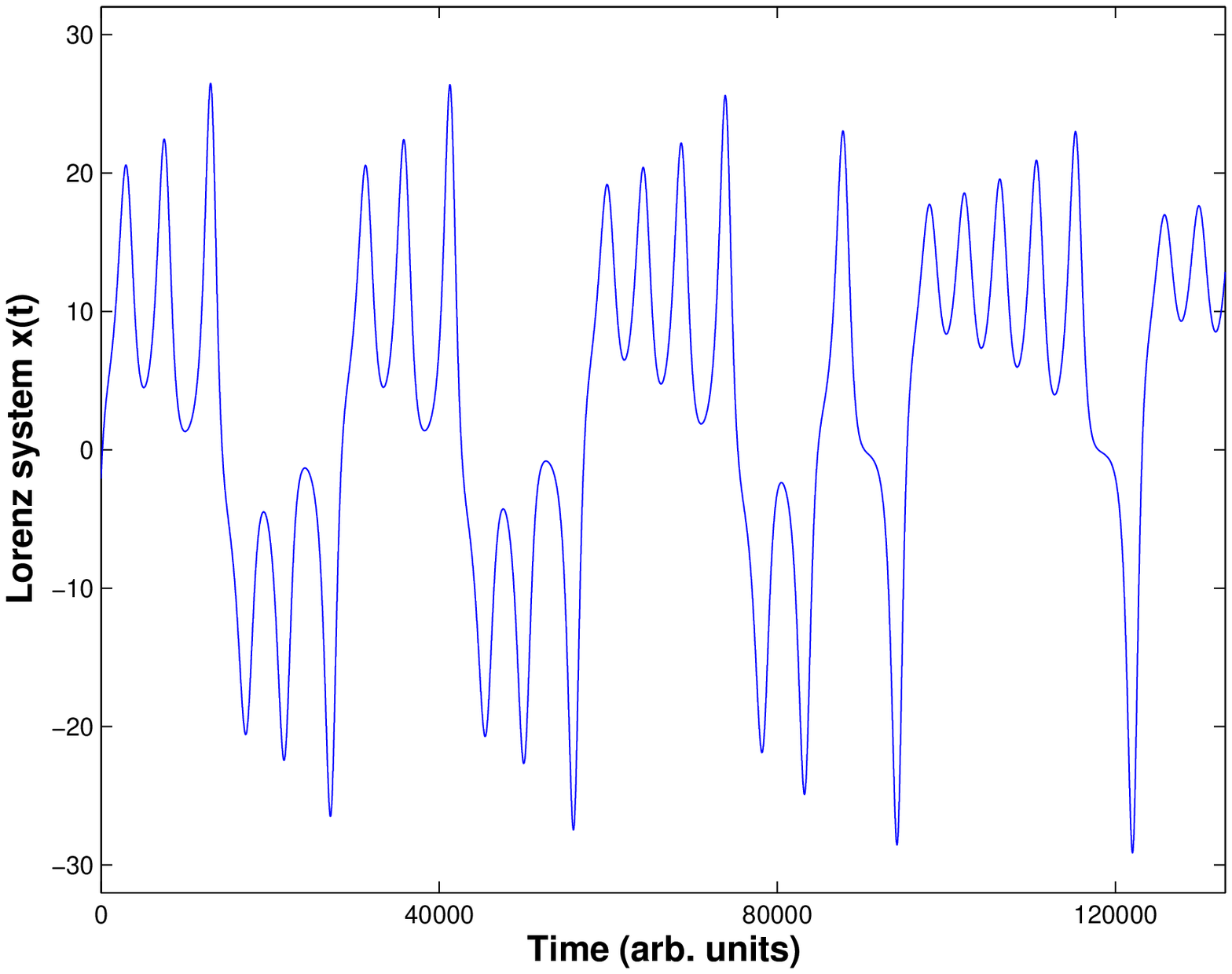}
\caption{Small segment of the $x(t)$ component of the Lorenz attractor
described in equations \ref{lorenzeqn} with $\kappa = 10^4$.}
\label{lorenz}
\end{center}
\end{figure}

We use a familiar example of a chaotic system, the Lorenz 
attractor~\cite{lorenz}, as the input signal to drive the R15 model. This is
a well studied system that exhibits chaotic dynamics and will be used here
as input to the R15 neuron model. The Lorenz attractor is defined by the
differential equations
\begin{eqnarray}\label{lorenzeqn}
\kappa \frac{dx(t)}{dt} & = & \sigma (y(t) - x(t)) \nonumber \\
\kappa \frac{dy(t)}{dt} & = & -x(t)z(t) + rx(t) - y(t) \\
\kappa \frac{dz(t)}{dt} & = & x(t)y(t) - bz(t) \nonumber
\end{eqnarray}
For the simulations presented in this paper the parameters
were chosen as $\sigma=16$, $r=45.92$ and $b=4$. The parameter
$\kappa$ is used to change the time scale. An example times series
of the $x(t)$ component of the Lorenz attractor is shown in
Figure \ref{lorenz}.

\subsection{Numerical Results}

An input signal $s(t) = I_{input}(t)$ is now formed from the $x(t)$
component of the Lorenz system. Our goal is to use observations of the
stimulus $I_{input}(t)$ and of the ISIs of the output signal $V_m(t)$ to
learn the dynamics of the R15 neuron model in the form of a local map in
phase space reconstructed from the observed ISIs. From this map we will
estimate the  input
$I_{input}(t)$ from new observations of the output ISIs.  

Our analog signal input is the $x(t)$ output of the Lorenz system, scaled
and offset to a proper range, and then input to the neuron as an external
current
\begin{equation}\label{scaledI}
I_{input}(t) = \mbox{Amp} (x(t) + x_0),
\end{equation}
where $\mbox{Amp}$ is the scaling constant and $x_0$ is the offset.  The
R15 equations are integrated~\cite{numrec} with this input signal and the
spike times $t_j$ from the membrane voltage are recorded simultaneously
with the value of the input current at that time $I_{input}(t_j)$.
Reconstruction of the neuron plus input phase space is done by creating
time delay vectors from the ISIs
\begin{equation} \label{isivec}
\y(j) = [isi_j, isi_{j+1},\ldots, isi_{j + (\de-1)\tau}]
\end{equation}
where 
\begin{equation} \label{isi} isi_j = t_j - t_{j-1}
\end{equation} 
For each of these vectors there is a corresponding value of
the input current which we chose to be at the midpoint time of the vector
\begin{equation} 
s(j) = I_{input}\left(t_{j + (\de-1)\tau/2}\right)
\end{equation}
In our work a total of 40000 spikes were collected. The first
30000 were used to create the training set vectors and the next 10000 were
used to examine our input estimation methods.  For each new output vector
constructed from new observed ISIs, $N_B$  nearest neighbors from the
training set were used to generate a local polynomial map $\y(j) \to
I^{estimated}_{input}(j)$. $N_B$ was chosen to be twice the number of free
parameters in the undetermined local coefficients $M_0, \M_1, \M_2, \ldots$.

We used the same three values of $I_{ext}$ -0.15, 0.1613, and 0.2031 employed
above in our
simulations.  We took $\mbox{Amp} = 0.001$, $\kappa = 10^4$, and
$x_0 = 43.5$  for all simulations unless stated otherwise. This very
small amplitude of the input current is much more of a challenge for
the input reconstruction than large amplitudes. When $\mbox{Amp}$ is
large, the neural activity is entrained by the input signal and
`recovering' the input merely requires looking at the output and
scaling it by a constant.  Further, the intrinsic spiking of the
neuron which is its important biological feature goes away when
$\mbox{Amp}$ is large.  The large value of $\kappa$ assures that the
spikes sample the analog signal $I_{input}(t)$ very well.

\begin{figure}[ht]
\begin{center}
\includegraphics[width=8cm,height=8cm,angle=0]{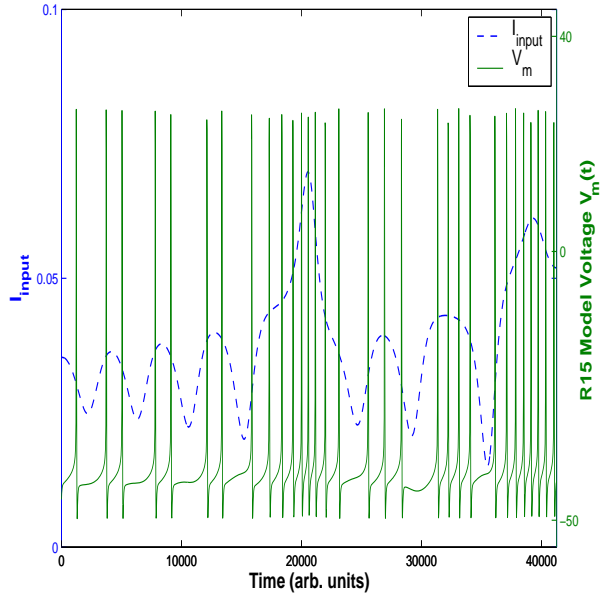}
\caption{A segment of the R15 neuron model output $V_m(t)$ shown along with
the scaled Lorenz system input current $I_{input}$.  Here
$I_{ext}=0.1613$, $\mbox{Amp} = 0.001$, and $\kappa = 10^4$. Note the
different scales for $I_{input}$ (shown on the left axis) and 
$V_m(t)$. (shown on the right axis)}
\label{spikes1613_001}
\end{center}
\end{figure}

For $I_{ext} = 0.1613$ we show a selection of both the input current
$I_{input}$ and the output membrane voltage $V_m(t)$ time series
in Figure~\ref{spikes1613_001}. The injected current substantially
changes the pattern of firing seen for the autonomous neuron.  Note
that the size of the input current is numerically about $10^{-3}$
of $V_m(t)$, yet the modulation of the ISIs due to this small input
is clearly visible in Figure \ref{spikes1613_001}.

\begin{figure}[ht]
\begin{center}
\includegraphics[width=8cm,height=8cm,angle=0]{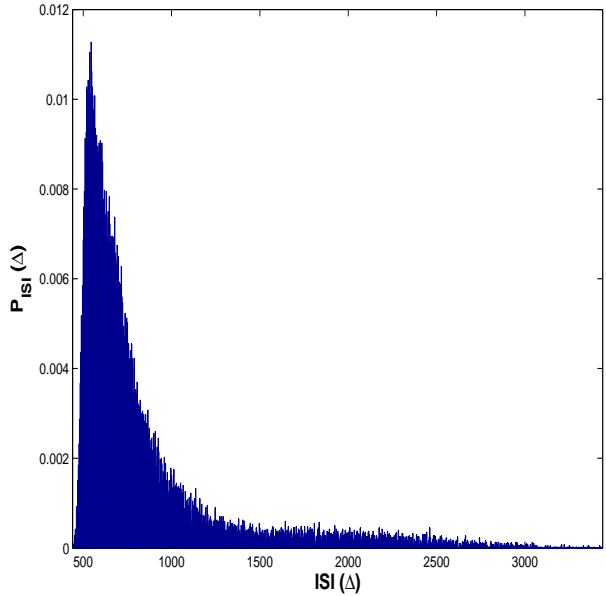}
\caption{$P_{ISI}(\Delta)$ for R15 model neuron output when a scaled $x(t)$
signal from the Lorenz system is presented with $I_{ext} = 0.1613$. 
The entropy of this distribution $H(\Delta) = 8.16$.}
\label{isidist1613_001}
\end{center}
\end{figure}

Using the ISIs of this time series we evaluated $P_{ISI}(\Delta)$ as
discussed above and from that the entropy $H(\Delta)$ associated with the
driven neuron. The ISI distribution, $P_{ISI}(\Delta)$, shown in 
Figure \ref{isidist1613_001}, has an entropy $H(\Delta) = 8.16$.
The effect of the input current has been to substantially narrow
the range of ISIs seen in $V_m(t)$. This can be seen by comparison
with Figure \ref{isidist1613}. 

\begin{figure}[ht] 
\begin{center}
\includegraphics[width=8cm,height=8cm,angle=0]{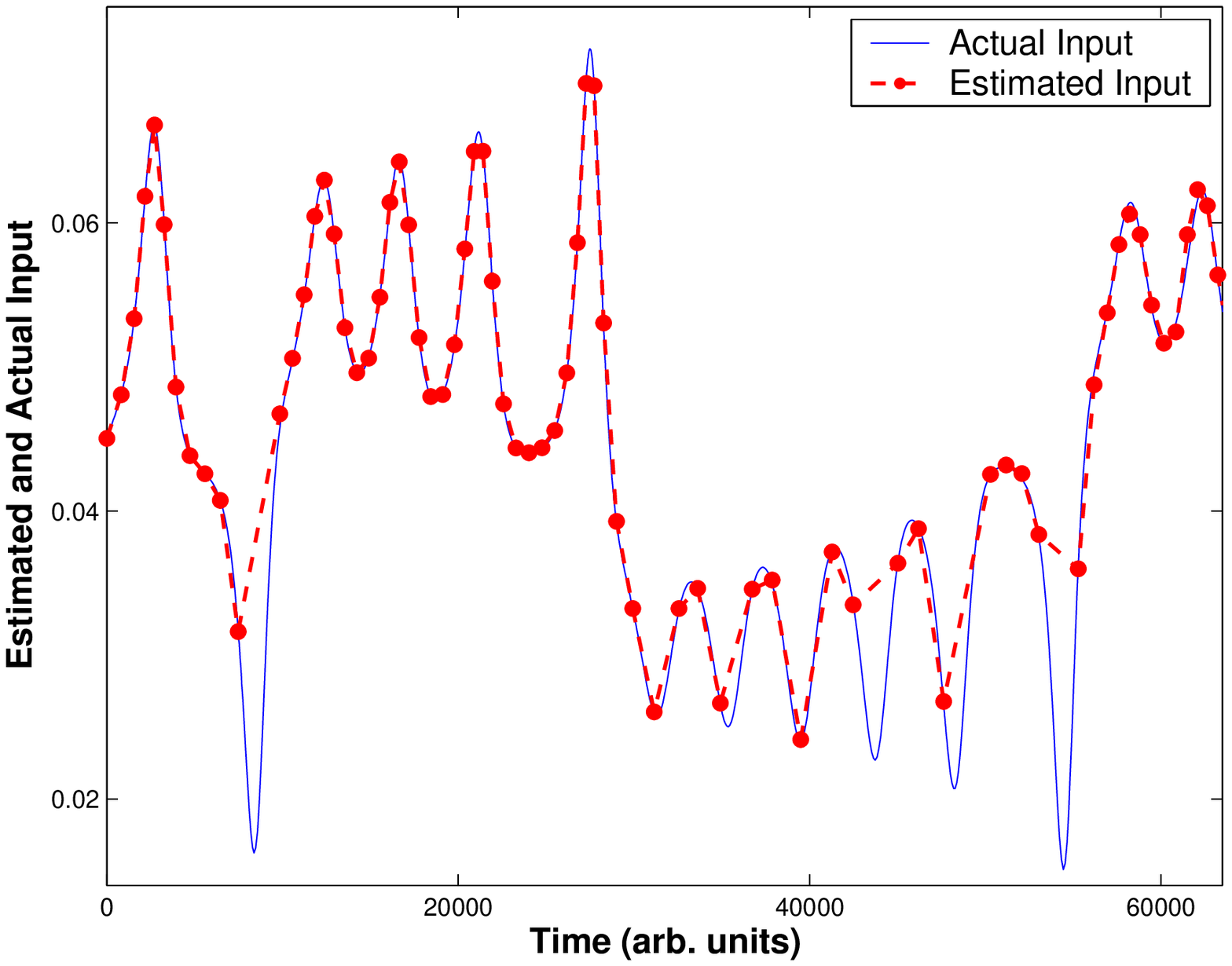}
\caption{ISI Reconstruction of the input Lorenz signal to an R15 neuron. The
solid line is the actual input to the neuron. The dots joined by dashed lines
are the ISI reconstructions.  The embedding dimension of the reconstruction
$\de$ is 4, the time delay $T$ is 1, $I_{ext} = 0.1613$, $\kappa$ is $10^4$,
and a linear map was used. The RMS error of the estimates over 10,000
estimations is $\sigma = 4.6\cdot10^{-4}$ and the maximum error is
about $0.01$.}
\label{r15_1613_001_1}
\end{center}
\end{figure}

Figure~\ref{r15_1613_001_1} shows an example of input signal reconstruction
which estimates $I_{input}$ using ISI vectors of the described in
Equation~\ref{isivec}.  We used a time delay $T = 1$, an embedding
dimension $\de = 7$, and a local linear map for $H(\y(j))$.
The RMS error over the 10,000 reconstructed values of the input
was $\sigma = 4.6\cdot10^{-4}$.   The input signal is only reconstructed 
at times at which the neuron spikes.  So each point is the reconstruction
curve in Figure \ref{r15_1613_001_1} corresponds to a spike in
$V_m(t)$.  Some features of the input are missed because no spikes
occur during that time, but otherwise the reconstruction is very
accurate. At places where the spike rate is high, interpolation seems
to fill the gaps between spikes.

Different values of embedding dimension, time delay, and map order
will lead to different reconstruction errors.  For example, low
embedding dimension may not unfold the dynamics and linear maps
may not be able to fit some neighborhoods to the input.  For the results
shown here, there is little difference in the RMS reconstruction error
if the embedding dimension is increased or quadratic maps are used instead
of linear maps.  This may not be true if lower embedding dimension is used.

\begin{figure}[ht]
\begin{center}
\includegraphics[width=8cm,height=8cm,angle=0]{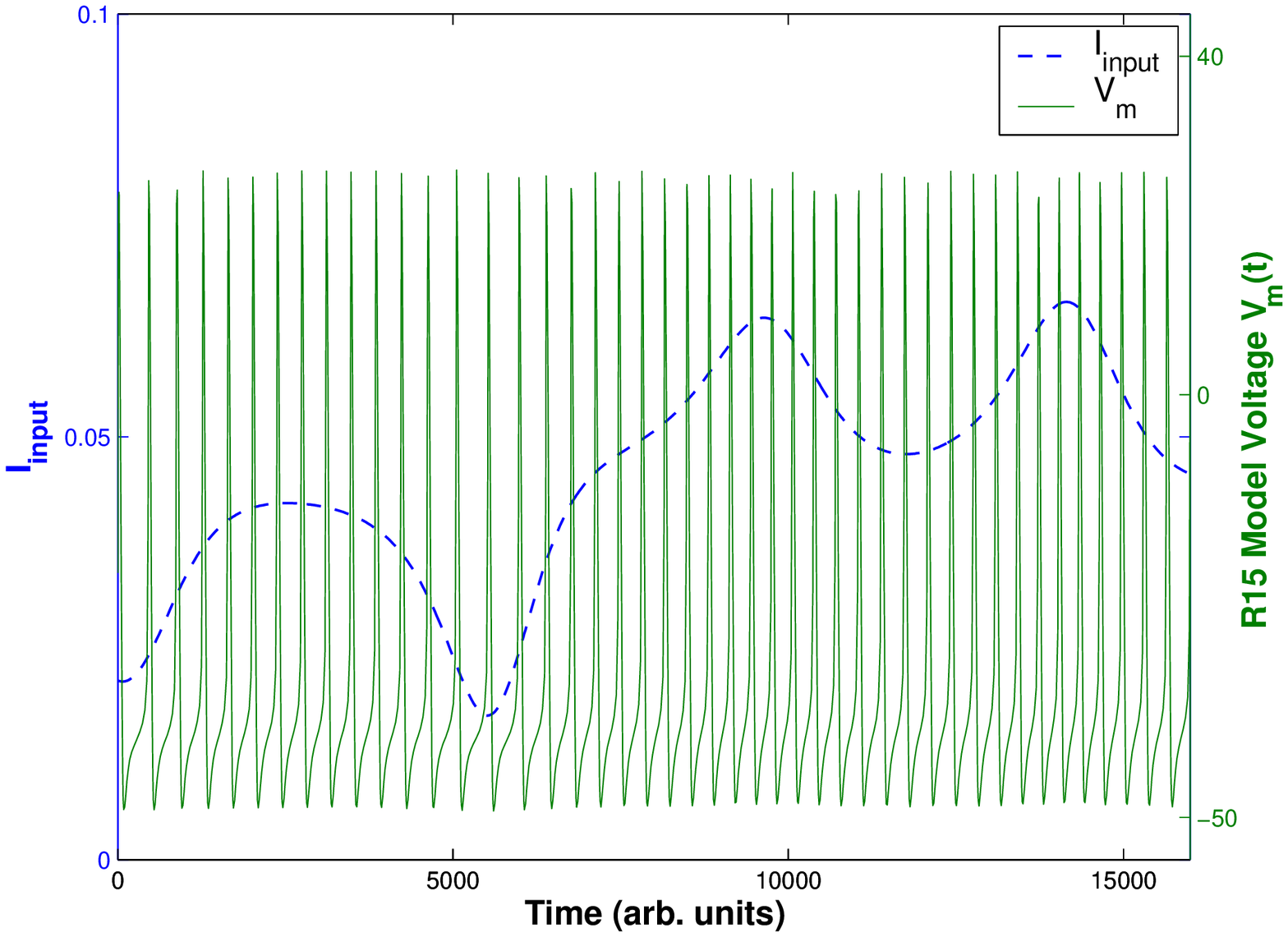}
\caption{A segment of the R15 neuron model output $V_m(t)$ shown along with
the scaled Lorenz system input current $I_{input}$.  Here
$I_{ext}=0.2031$,$\mbox{Amp} = 0.001$, and $\kappa = 10^4$. Note the
different scales for $I_{input}$ and $V_m(t)$.}
\label{spikes2011_001}
\end{center}
\end{figure}

\begin{figure}[ht]
\begin{center}
\includegraphics[width=8cm,height=8cm,angle=0]{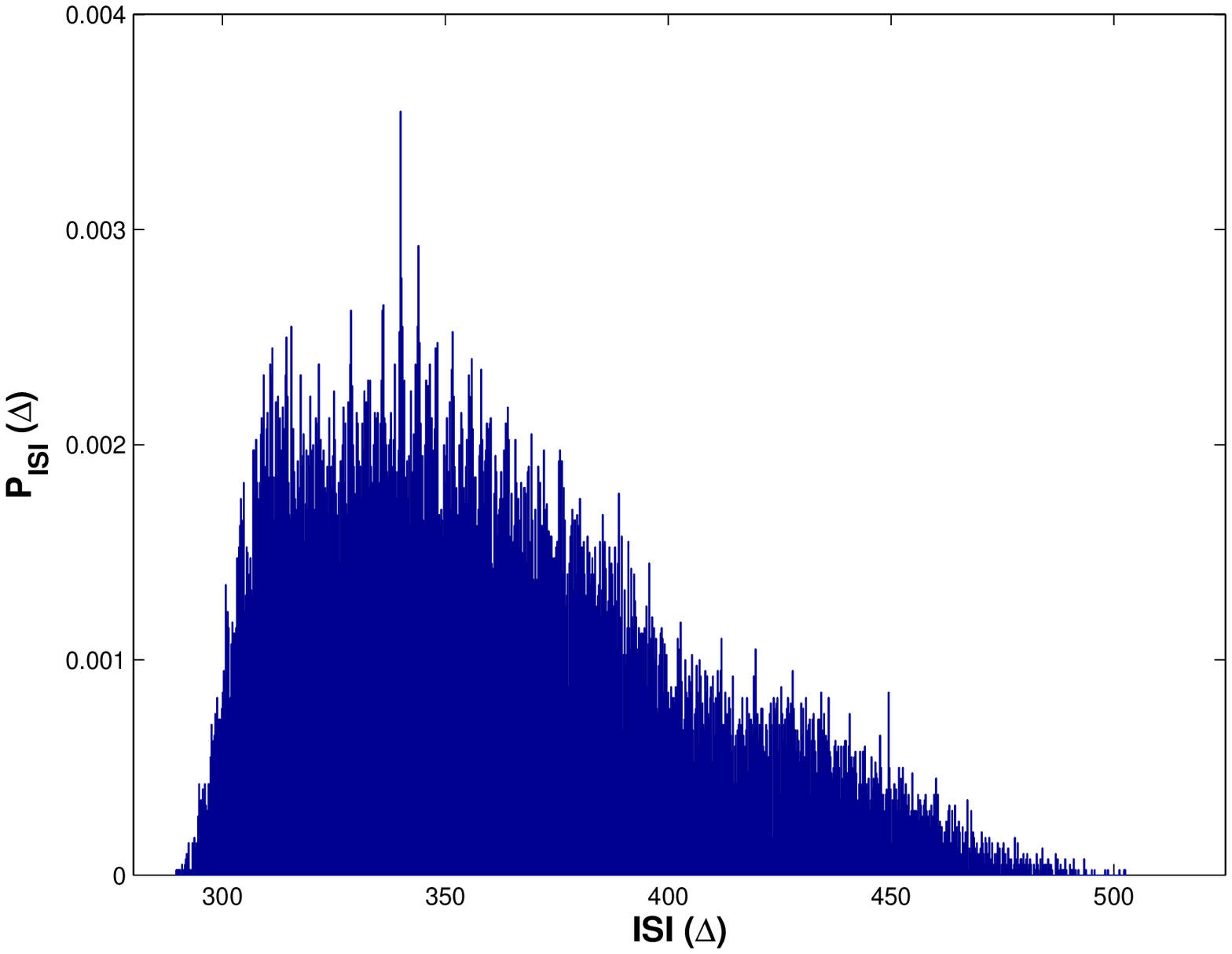}
\caption{$P_{ISI}(\Delta)$ for R15 model neuron output when a scaled $x(t)$
signal from the Lorenz system is presented with $I_{ext} = 0.2031$. 
The entropy of this distribution $H(\Delta) = 9.5$.}
\label{isidist2011_001}
\end{center}
\end{figure}

\begin{figure}[ht] 
\begin{center}
\includegraphics[width=8cm,height=8cm,angle=0]{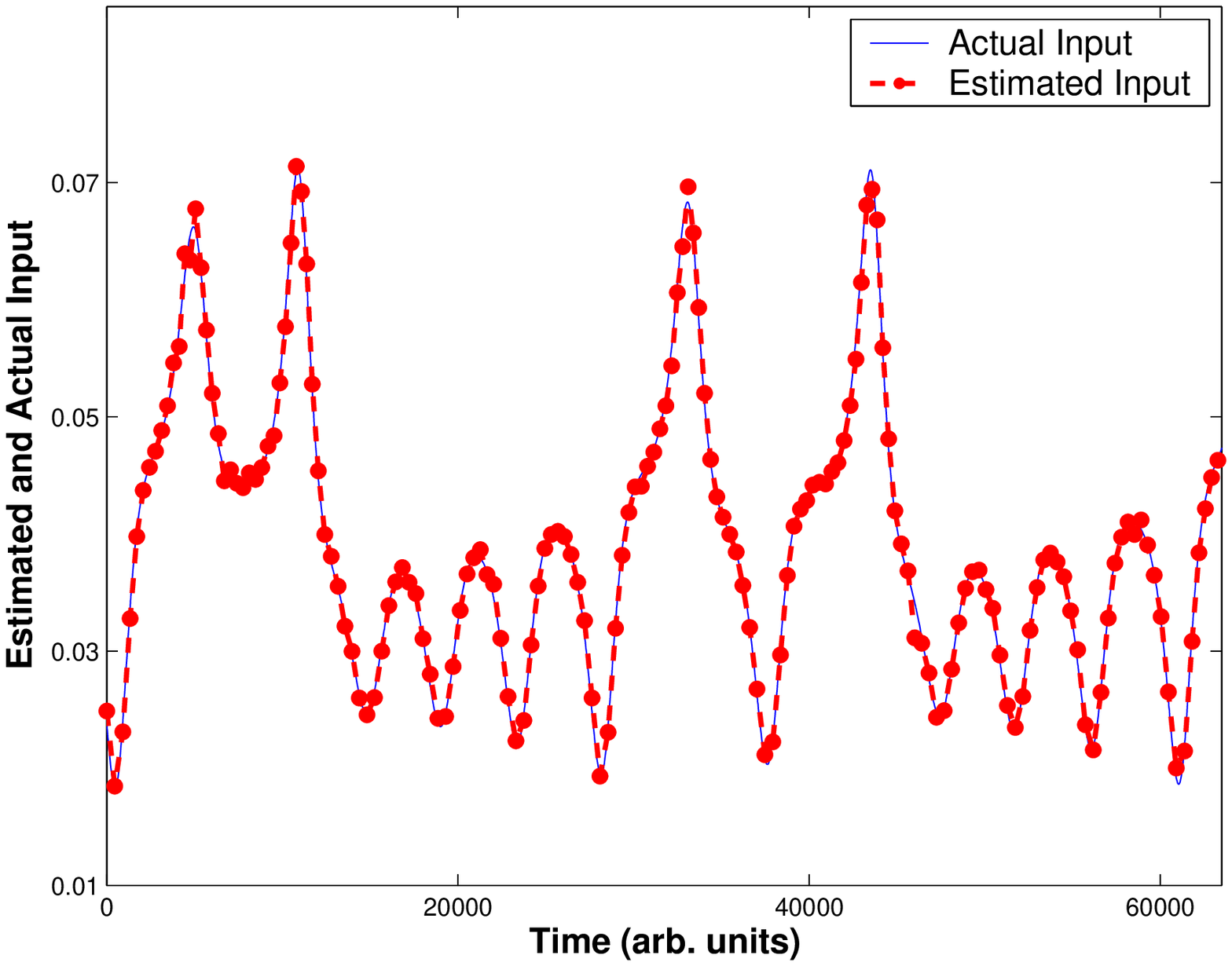}
\caption{ISI Reconstruction of the input Lorenz signal to an R15 neuron. The
solid line is the actual input to the neuron. The dots joined by dashed lines
are the ISI reconstructions.  The embedding dimension of the reconstruction
$\de$ is 2, the time delay $T$ is 1, $I_{ext} = 0.2031$, $\kappa$ is $10^4$,
and a linear map was used. The RMS error of the estimates over 10,000
estimations is $\sigma = 6.1\cdot10^{-4}$ and the maximum error is
about $0.007$.}
\label{r15_2011_001_1}
\end{center}
\end{figure}

The previous example probed the response of a chaotic neural oscillation
to a chaotic signal.  With $I_{ext} = 0.2031$ the neuron is in a
periodic spiking regime and the input modulates the instantaneous
firing rate of the neuron.  A sample of the input current and
membrane voltage is shown in Figure~\ref{spikes2011_001}.
The distribution of ISIs, $P_{ISI}(\Delta)$, shown in
Figure~\ref{isidist2011_001} and has an entropy $H(\Delta) = 9.5$.
The effect of the input current is to substantially broaden the range
of ISIs and increase its entropy as compared to the nearly periodic firing
of the autonomous neuron with $I_{ext}=0.2031$.  The high spiking
rate and close relationship between input current amplitude and ISI
lead to very accurate reconstructions using low dimensional embeddings.
A sample of the reconstruction using $\de = 2$ and $T = 1$ is shown in
Figure~\ref{r15_2011_001_1}.  The RMS reconstruction error of
$\sigma = 6.1\cdot10^{-4}$ with a maximum error of $0.007$.

\begin{figure}[ht]
\begin{center}
\includegraphics[width=8cm,height=8cm,angle=0]{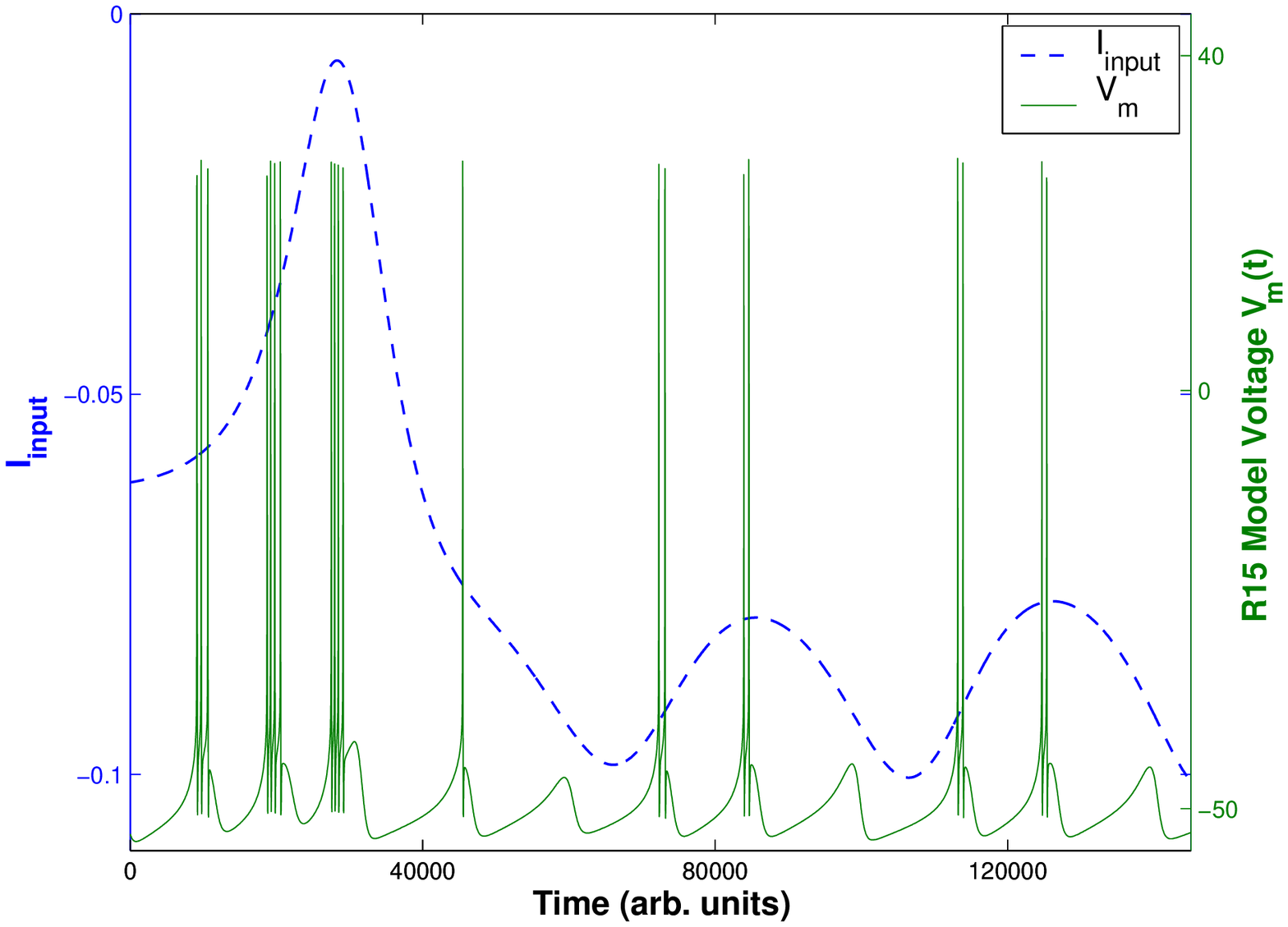}
\caption{A segment of the R15 neuron model output $V_m(t)$ shown along with
the scaled Lorenz system input current $I_{input}$.  Here
$I_{ext}=-0.15$,$\mbox{Amp} = 0.002$, and $\kappa = 2\cdot10^5$. Note the
different scales for $I_{input}$ and $V_m(t)$.}
\label{spikesn1500_002}
\end{center}
\end{figure}

\begin{figure}[ht]
\begin{center}
\includegraphics[width=8cm,height=8cm,angle=0]{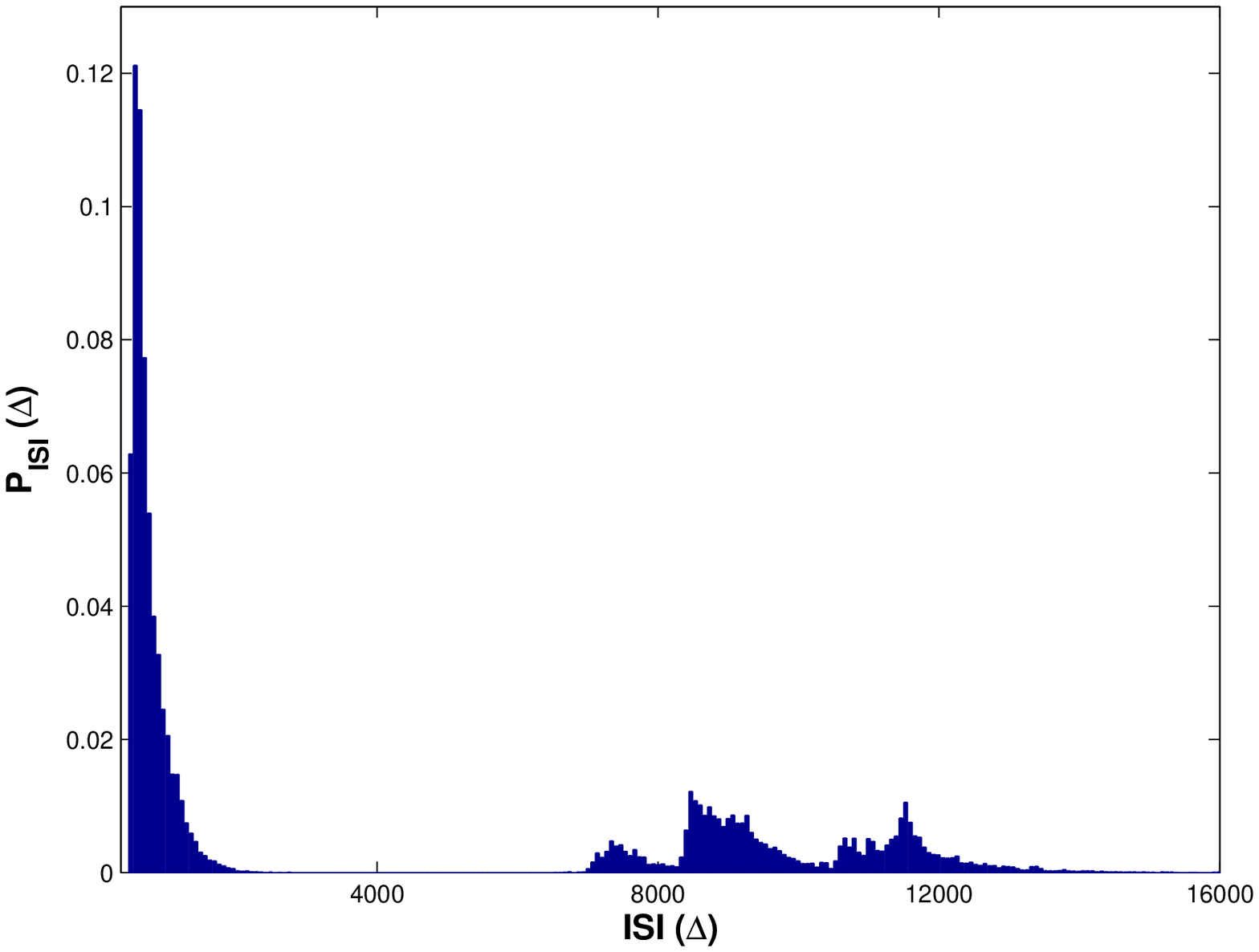}
\caption{$P_{ISI}(\Delta)$ for R15 model neuron output when a scaled $x(t)$
signal from the Lorenz system is presented with $I_{ext} = -0.15$. 
The entropy of this distribution $H(\Delta) = 5.3$.}
\label{isidistn1500_002}
\end{center}
\end{figure}

\begin{figure}[ht] 
\begin{center}
\includegraphics[width=8cm,height=8cm,angle=0]{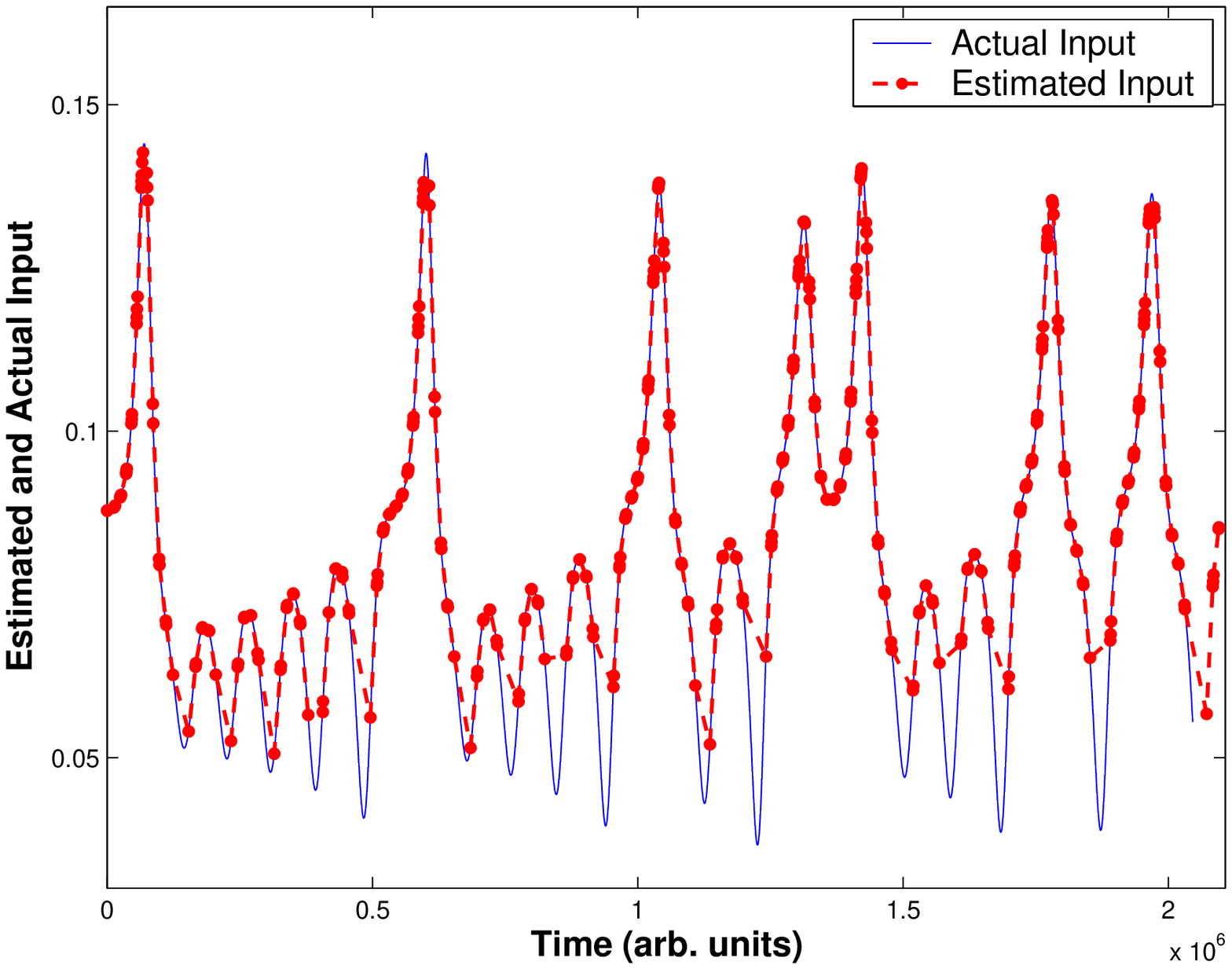}
\caption{ISI Reconstruction of the input Lorenz signal to an R15 neuron. The
solid line is the actual input to the neuron. The dots joined by dashed lines
are the ISI reconstructions.  The embedding dimension of the 
reconstruction $\de$ is 7, the time delay $T$ is 1, $I_{ext} = -0.15$,
$\kappa$ is $2\cdot10^5$, and a linear map was used. The RMS error of
the estimates over 10,000 estimations is $\sigma = 0.0094$ and the
maximum error is about $0.03$.}
\label{r15_n1500_002}
\end{center}
\end{figure}

In a final example we show the reconstruction when the neuron is being driven
with an input current below the threshold for spikes.  With $I_{ext} = -0.15$,
the autonomous R15 neuron will remain at an equilibrium level and not produce
spikes.  A Lorenz input injected into the neuron with $\mbox{Amp} = 0.002$
and $x_0 = 43.5$ is large enough to cause the neuron to spike.
Figure~\ref{spikesn1500_002} shows a sample of the membrane voltage time
series along with the corresponding input current.  Since the spiking rate
of the neuron is much lower than before, $\kappa$ is increased to $2\cdot10^5$.
This slows down the dynamics of the Lorenz input relative to the neuron
dynamics.  Spikes occur during increasing portions of the input current
and are absent for low values of input current.  Figure~\ref{isidistn1500_002}
shows the distribution of ISIs which has an entropy $H(\Delta)=5.3$.
The low spiking rate shows up in the distribution in the form large numbers
of long ISI.  For the reconstruction of the input larger embedding
dimensions were needed.  An sample of the reconstruction is shown
in Figure~\ref{r15_n1500_002} using $\de = 7$ and $T = 1$. For this fit
the RMS reconstruction error $\sigma = 0.0094$ with a maximum error
of $0.03$.  These errors are noticeably higher than the previous two examples.

The accuracy of the reconstruction method depends on a high spiking rate
in the neuron relative to the time scale of the input signal, since only
one reconstructed input value is generated for each spike.  If the spiking
rate of the neuron is low relative to the time scales of the input signal,
then the neuron will undersample the input signal
and miss many of its features. This limitation can be demonstrated
by decreasing the time scale parameter $\kappa$, thereby speeding up the
dynamics of the input. During the longer ISIs the input current can change
by large amounts. Though the reconstruction undersamples the input, but
interpolation can fill in some of the gaps. As $\kappa$ is increased
further the reconstruction will further degrade.

\section{Discussion}

In previous research on the encoding of chaotic attractors in spikes trains,
the spike trains were produced by nonlinear transformations of chaotic input
signals.  Threshold crossing neuron models have been used, which generate
the spike times at upward crossings of a threshold. This is equivalent to a
Poincare section of the input signal.  Also integrate and fire neurons have
been studied, which integrate the input signal and fire a spike when it
crosses a threshold, after which the integral is reset to zero.  Both of
these models have no intrinsic complex dynamics; they can not produce
entropy autonomously.  All of the complex behavior is in the input signal.
Even though the attractor of a chaotic input can be reconstructed from the
ISIs, these models do not account for the complex behavior of real neurons. 
The input reconstruction method we have presented here allows for complex
intrinsic dynamics of the neuron.  We have shown that the local polynomial
representations of input/output relations realized in reconstructed phase
space can extract the chaotic input from the complex interaction between the
input signal and neuron dynamics.

Other experimental works have used linear kernels to map the spike train
into the input.  They have shown that the precise timing of individual
spikes can encode a lot of information about the input~\cite{spikes}.  And
the precise relative timing between two spikes can carry even more
information than their individual timings combined~\cite{synergy}.  These
results may be pointing toward a state space representation since the time
delay embedding vectors used here take into account both the precise spike
timing and the recent history of ISIs.   From a dynamical systems
perspective this is important because the state of the system at the time of
the input will affect its response.  This is a factor that linear kernels do
not take into account.

The advantage of using local representations of input/output relations in
reconstructed state space lies primarily in the insight it may provide about
the underlying dynamics of the neural transformation process mapping analog
environmental signals into spike trains. The goal of the work presented here
is not primarily to show we can accurately recover analog input signals from
the ISIs of spike output from neurons, though that is important to
demonstrate. The main goal is to provide clues on how one can now model the
neural circuitry which transforms these analog signals. The main piece of
information in the work presented here lies in the size of the reconstructed
space $\de$ which tells us something about the required dimension of the
neural circuit. Here we see that a low dimension can give excellent results
indicating that the complexity of the neural circuit is not fully utilized
in the transformation to spikes. Another suggestion of this is in the
entropy of the input and output signals. In the case where 
$I_{ext} = 0.1613$ the entropy of the analog input is $11.8$ while the entropy
of the ISI distribution of the output is $8.16$. When $I_{ext} = 0.2031$ the
output entropy is $9.5$. This suggests, especially in the case of the larger
current, that the signal into R15 neuron model acts primarily as a
modulation on the ISI distribution. This modulation may be substantial, as
in the case when $I_{ext} = 0.1613$ but reading the modulated signal does
not require complex methods.

Our final example  took $I_{ext} = -0.15$ at which value the undriven neuron
has $V_m(t) = \mbox{constant}$, so it is below threhold for production
of action potentials. In this case the introduction of the stimulus drove
the neuron above this threshold and produced a spike train which could be
accurately reconstructed.  This example is
relevant to the behavior of biological neurons which act as sensors for various
quantities: visiual stimuli, chemical stimuli (olfaction), etc.
In the study of biological sensory 
systems~\cite{tumer} the neural circuitry is quiet in the absence of
input signals, yet as we now see the methods are equally valid and accurate.

\clearpage

\section*{Acknowledgements} This work was partially  supported by the U.S.
Department of Energy, Office of Basic Energy Sciences, Division of
Engineering and Geosciences, under Grants No. DE-FG03-90ER14138 and No.
DE-FG03-96ER14592, by a grant from the National Science Foundation, NSF
PHY0097134, by a grant from the Army Research Office, DAAD19-01-1-0026, by a
grant from the Office of Naval Research, N00014-00-1-0181, and by a grant
from the National Institutes of Health, NIH R01 NS40110-01A2. ET acknowledges
support from NSF Traineeship DGE 9987614.

\clearpage

\bibliographystyle{plain}
\bibliography{main}

\end{document}